\documentclass[aps, pra, a4paper, showpacs, twocolumn, english, 10pt, nofootinbib]{revtex4-1}
\usepackage[T1]{fontenc}
\usepackage{babel}

\usepackage{bbm, amsthm, bm, textcomp, nicefrac, geometry, ragged2e}
\usepackage[dvipsnames]{xcolor}
\usepackage{float}
\usepackage[bbgreekl]{mathbbol}
\usepackage{graphicx, epstopdf, color, verbatim, enumitem, ulem}
\usepackage{pbox}
\usepackage{mathrsfs}
\usepackage{amssymb}
\usepackage{amsfonts}
\usepackage{mathtools}
\usepackage{stackrel}
\usepackage[thinlines]{easytable}
\usepackage{amsmath}
\usepackage{makecell}
\usepackage{array}
\usepackage{multirow}

\mathtoolsset{showonlyrefs}

\usepackage{tikz}
\usetikzlibrary{shapes,arrows, positioning, calc, quantikz,angles, quotes}  
\makeatletter

\usepackage[caption=false]{subfig}
\addto\captionsenglish{}
\usepackage{hyperref}

\hypersetup{
    colorlinks = true,
    linkcolor = BrickRed,
    citecolor = BrickRed,
    filecolor = Black,     
    urlcolor = Black,
}
\newcommand*{\rom}[1]{\expandafter\@slowromancap\romannumeral #1@}
\DeclareMathAlphabet{\mathbbold}{U}{bbold}{m}{n}
\DeclareMathSymbol{\shortminus}{\mathbin}{AMSa}{"39}

\renewcommand\bra[1]{{\langle{#1}|}}
\renewcommand\ket[1]{{|{#1}\rangle}}
\begin{document}

\title{Improved Quantum Sensing by Spectral Design} %Adaptive quantum sensing
\author{Paul Aigner}
\email[Corresponding author: ]{paul.aigner@uibk.ac.at}
\affiliation{Institut f\"ur Theoretische Physik, Universit\"at Innsbruck, Technikerstra{\ss}e 21a, 6020 Innsbruck, Austria}
\author{Wolfgang Dür}
\affiliation{Institut f\"ur Theoretische Physik, Universit\"at Innsbruck, Technikerstra{\ss}e 21a, 6020 Innsbruck, Austria}

\date{\today}

%% Abstract %%
\begin{abstract}
We investigate how unitary control can improve parameter estimation by designing the effective spectrum of the imprinting Hamiltonian. We show that, for commuting Hamiltonians, the general problem of spectral manipulation via unitary control simplifies to a finite sequence of elementary switching operations. Furthermore, we demonstrate that any desired relative spacing of energy levels can be achieved, although this may come at the cost of a reduced spectral range. We also show that any modified spectrum can be expressed as a convex combination of the original eigenvalues, with the convex weights forming a bi-stochastic matrix. Through several single-parameter estimation examples, we demonstrate that our spectral engineering method substantially enhances estimation accuracy.

\end{abstract}

%% Title %%
\maketitle

%% Body %%
\section{Introduction} \label{Sec.Introduction}
Quantum sensing exploits quantum systems to measure physical parameters with high precision. A key challenge in quantum metrology is designing sensing protocols adaptable to diverse problem settings and capable of robust performance across various scenarios. Standard metrology fixes the parameter-imprinting process, e.g. the unitary evolution under a specified Hamiltonian, and then optimizes the probe state, the measurement, and the estimator. Here we treat the signal imprinting process itself as another control knob. By applying unitary control we reshape the system’s evolution and, in effect, redesign the sensor on the fly. Static configurations are powerful in narrowly defined settings, but may not be optimal for a wide range of sensing tasks.  Consequently, adaptive quantum sensing protocols, which dynamically adjust their operational parameters in response to the sensing task, have become important tools for improving estimation accuracy and robustness \cite{Pang2017,PhysRevX.7.041009,PhysRevX.8.021059,sekatski2017quantum,PhysRevA.84.052315,PhysRevLett.107.233601,PhysRevLett.89.133602,PhysRevLett.110.220501}.

In this paper, we introduce an adaptive quantum sensor capable of dynamically modifying the effective energy spectrum of a Hamiltonian through unitary control. Unlike previous Bayesian adaptive estimation strategies that primarily focus on iterative parameter updates \cite{PhysRevA.84.052315,PhysRevLett.107.233601,PhysRevLett.89.133602,PhysRevLett.110.220501}, our technique uses spectral manipulation, which systematically reshapes the underlying Hamiltonian dynamics to optimize estimation precision. Particularly, we can manipulate and optimize the signal imprinting process, thereby improving the performance of the sensor.

Our approach, termed the switching method, reduces the general problem of spectral adaptation, for commuting Hamiltonians, via unitary control to a finite series of elementary so-called switching operations. We demonstrate that any achievable spectral transformation can be represented as a convex combination of the Hamiltonian’s original eigenvalues, where the convex weights form a bi-stochastic matrix. This offers a systematic framework for spectral engineering distinct from other quantum control methods such as dynamical decoupling, which typically focus on noise suppression or fidelity enhancement rather than explicit spectral adaptation \cite{PhysRevX.7.041009,PhysRevX.8.021059,sekatski2017quantum}. The applicability of our method is restricted to the Bayesian parameter estimation regime, where the exact structure of the spectrum, such as the total number of non-degenerate levels, plays a role \cite{wolk2020noisy}. In contrast, within the Fisher information regime, our method offers no fundamental advantage, since it cannot increase the spectral range, which solely determines the optimal estimation performance \cite{10.1116/1.5119961}.

We showcase our method's improvements in sensing precision in many distinct scenarios through various examples in single-parameter estimation:
\renewcommand{\theenumi}{\roman{enumi}}%
\begin{enumerate}
    \item Degenerate spectra
    \item Physically motivated non-linear spectra
    \item Different prior distributions
\end{enumerate}
Our method can obtain clear advantages in settings with structured priors, as well as for Gaussian priors if the initial spectrum is not linear or when degeneracies can be exploited. In contrast, for Gaussian priors the linear spectrum is already almost optimal, and additional optimization yields only marginal gains. We therefore emphasize that the contribution of our approach lies not in universally improving sensing performance, but in offering substantial benefits in regimes where prior information or distinct spectral features can be utilized.

The structure of the paper is as follows. In Sec.~\ref{Sec.Method} we introduce the so-called switching method for adapting the effective spectrum of a Hamiltonian for the imprinting of a to be estimated parameter. We show that the effective spectrum is given via a convex combination of the original spectrum and that different spectral engineering is not possible even with arbitrary unitary control. In addition, we demonstrate the feasibility of achieving arbitrary relative energy spacings, while highlighting that such spectral adaptation may incur a reduction in the spectral range, which in the worst case can be reduced by a factor of $n-1$, where $n$ denotes the number of levels. Next, in Sec.~\ref{Sec:single_parameter_bayesian_estimation} we give a short review of the concepts of single parameter Bayesian frequency (phase) estimation. This is followed, in Sec.~\ref{Sec:Application of method}, where we go through different physical set-ups, ranging from degenerate spectra to non-linear spectra and from flat prior to highly structured priors, and show via examples the utility of the method for increasing the estimation performance.

Finally, in Sec.~\ref{Sec: Conclusion} we give a concise overview of the results and give an outlook for future research directions. 

\section{Method}\label{Sec.Method}
\subsection{Set-up}

\begin{figure}
\includegraphics{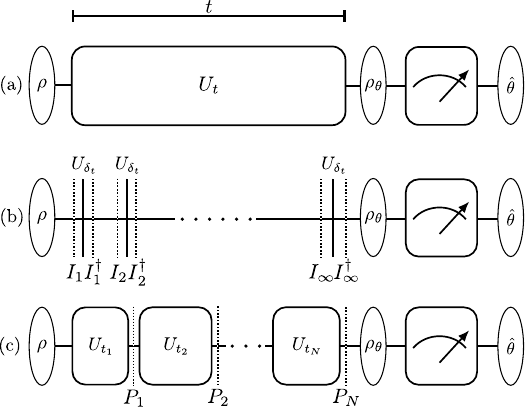}
   \caption{
Illustration of quantum sensing protocols with and without unitary control.  
(a) Standard sensing process using unitary imprinting of parameter $\theta$.  
(b) Sensing with general unitary control implemented via an infinite sequence of intermediate control unitary operations $I_i$.  
(c) Sensing using the switching method, based on a finite sequence of permutation operations $P_i$.
}
    \label{fig:circuit}
\end{figure}
In quantum metrology, a fundamental task is to estimate an unknown parameter that appears in the dynamics of a quantum system. A common and important scenario involves single-parameter estimation \cite{10.1116/1.5119961}, where the parameter governs the evolution of a known quantum system. To formalize this, we consider a Hamiltonian of the form
\begin{equation} \label{equ:initial:ham}
H = \omega G = \omega \sum_{i=0}^{n-1} \lambda_i \ket{i} \bra{i},
\end{equation}
where $\omega$ is the unknown parameter we aim to estimate, such as the strength of a magnetic field, and $G$ is a known Hermitian operator called the generator of the dynamics, with a known spectrum $\sigma(G) = \{\lambda_i\}$.

This structure is typical in physical systems where the parameter of interest couples to the system through a fixed generator. For example, in magnetometry, a spin system evolves under the influence of a magnetic field whose strength $\omega$ we want to determine. The generator $G$ in this case corresponds to a spin observable aligned with the magnetic field.

To extract information about $\omega$, we prepare an initial quantum state $\rho$, called the probe state, and let it evolve under the unitary dynamics generated by the Hamiltonian. The evolution is governed by the unitary operator
\begin{equation}
U_t = e^{-i t \omega G},
\end{equation}
which depends on both time $t$ and the unknown parameter $\omega$. After the unitary imprinting process, the state becomes
\begin{equation}
\rho_{\omega,t} = U_t \rho U_t^\dag,
\end{equation}
with $\omega$ now encoded in the quantum state. For the special case where the initial state is a pure state $\ket{\Psi}$, the evolved state takes the form
\begin{equation} \label{equ:transf:pure}
\ket{\Psi}_{\omega,t} = U_t \ket{\Psi} = \sum_{i=0}^{n-1} c_i e^{-i t \omega \lambda_i} \ket{i},
\end{equation}
where the coefficients $c_i$ come from the decomposition of $\ket{\Psi}$ in the eigenbasis $\{\ket{i}\}$ of the generator $G$. This transformation clearly shows how the unknown parameter $\omega$ affects the phase evolution of the quantum state, making it possible to infer $\omega$ through suitable measurements.

\subsection{Switching Method}
In a standard sensing protocol for a given physical set-up described by the Hamiltonian $H$ we first prepare the probe state, then imprint an parameter via unitary dynamics, followed by the measurement of the state, and in the end we estimate the parameter, as illustrated in Fig.~\ref{fig:circuit}(a). The choice of probe states and measurement, together with using an appropriate estimator is usually used to optimize the sensing process.
Here we additionally permit during the imprinting process intermediate unitary operations, as illustrated in Fig.~\ref{fig:circuit}(c). Specifically, we consider so-called \textit{switching operations} defined as
\begin{equation} S_{ij} = \ket{i}\bra{j} + \ket{j}\bra{i} + \sum_{k \neq i \lor j} \ket{k}\bra{k}, \end{equation}
applied at intermediate times 
$t_i \in [0,t]$ during the evolution. Equivalently, one can consider permutation unitary matrices, which act as permutations on the computational basis, since any permutation can be decomposed into the permutations given by the switching operations. 

The intermediate unitary operations we consider can be understood as a form of quantum control in which a finite number of very short control pulses from a specific gate family are applied during the interrogation time. This is in contrast to full and fast quantum control where an arbitrary number of arbitrary operations are allowed \cite{Sekatski2017quantummetrology}. It also differs from the soft unitary control setting where a continuously modulated control Hamiltonian is switched on at the beginning of the interrogation and switched off at the end \cite{PhysRevLett.121.050402}. In both quantum control and digital Hamiltonian simulation control methods are usually employed either to suppress noise \cite{PhysRevLett.82.2417,PhysRevLett.90.037901,PhysRevLett.134.120802} or to generate effective interactions \cite{RevModPhys.86.153, PhysRevLett.79.2586, PhysRevA.65.040301}. Our approach instead uses quantum control to reshape the imprinting dynamics with the aim of enhancing sensing performance.

These intermediate operations can alter the Hamiltonian’s effective energy spectrum, and hence the imprinting of the parameter on the sensing state. To illustrate, consider the action of $S_{ij}$ at time $t_0$ on the evolved state $\ket{\Psi}_{\omega,t_0}$ 

\begin{equation} \begin{aligned} S_{ij} \ket{\Psi}_{\omega, t_0} =& \ c_i e^{-i t_0 \lambda_i \omega} \ket{j} + c_j e^{-i t_0 \lambda_j \omega} \ket{i} \\ &+ \sum_{k \neq i \lor j} c_k e^{-i t_0 \lambda_k \omega} \ket{k}. \end{aligned} \end{equation}

After further unitary evolution and reapplying 
$S_{ij}$ at time $t$, the state becomes
\begin{equation} \begin{aligned} S_{ij} U_{t - t_0} S_{ij} \ket{\Psi}_{\omega, t_0} =& \ c_i e^{-i \omega \left[ r_0 \lambda_i + (1 - r_0) \lambda_j \right]} \ket{j} \\ &+ c_j e^{-i \omega \left[ r_0 \lambda_j + (1 - r_0) \lambda_i \right]} \ket{i} \\ &+ \sum_{k \neq i \lor j} c_k e^{-i \lambda_k \omega t} \ket{k}, \end{aligned} \end{equation} 
where $r_0=t_0/t$. 

We have thus effectively modified the eigenvalues 
$\lambda_i$ and $\lambda_j$, replacing them with effective eigenvalues

\begin{equation}
    \lambda^{(\text{eff})}_i=r_0 \lambda_i + (1 - r_0)\lambda_j, \ \lambda^{(\text{eff})}_j=r_0\lambda_j + (1 - r_0)\lambda_i,
\end{equation}
which are convex combinations of the original eigenvalues. Consequently, the controlled evolution described above can be interpreted as a free evolution under an effective Hamiltonian $H_{\text{eff}}$, whose spectrum incorporates these modified eigenvalues.
In the general case, with an arbitrary number of switching operations, the effective spectrum takes the form
\begin{equation}
    \lambda_i^{(\text{eff})}=\sum_j R_{ij} \lambda_j, \ \sum_{i} R_{ij}= \sum_{j} R_{ij}=1, \ R_{ij} \geq 0,
\end{equation}
where $R_{ij}$ denotes the relative time that energy level labeled by $j$ contributes to the effective level labeled by $i$. The convex weights $R_{ij}$ form a bi-stochastic matrix, i.e. a matrix with positive entries which rows and columns sum to one. As the effective spectrum is a convex combination of the original spectrum the effective spectral range, i.e. the difference between the maximum and the minimum eigenvalue, is in general reduced. Although the reduction of the spectral range is generally detrimental to sensing performance, it can often be compensated by the advantages gained from modifying the level structure, as demonstrated in several examples in Sec.~\ref{Sec:Application of method}. 

\subsubsection{Analytical solution}
Here we show how to realize any desired relative spectrum, i.e., to engineer arbitrary energy spacings relative to the overall spectral range. This can be expressed using a set of \textit{target ratios} \( t_i \in [0, 1] \), defined by
\begin{equation} \label{equ:7}
    t_i=\frac{\lambda^{(\text{eff})}_i - \lambda^{(\text{eff})}_{\min}}{\lambda^{(\text{eff})}_{\max} - \lambda^{(\text{eff})}_{\min}} ,
\end{equation}
where \( \lambda^{(\text{eff})}_{\min} \) and \( \lambda^{(\text{eff})}_{\max} \) denote the minimum and maximum effective eigenvalues, respectively. We collect the target ratios in a so-called target vector $\boldsymbol{t}=(t_0,t_1,\dots,t_{n-1})$.

Rewriting this condition as a constraint on the relative times \( R_{ij} \), and assuming without loss of generality that \( \lambda^{(\text{eff})}_{\min} = \lambda^{(\text{eff})}_0 \) and \( \lambda^{(\text{eff})}_{\max} = \lambda^{(\text{eff})}_{n-1} \), we obtain
\begin{equation} \label{equ:8}
\begin{aligned}
 \sum_{j=0}^{n-1} \lambda_j \left[ R_{ij} - \left( (1 - t_i) R_{0j} + t_i R_{(n-1)j} \right) \right] = 0, \\
  \sum_{i} R_{ij}= \sum_{j} R_{ij}=1, \quad 
    R_{ij} \geq 0.
\end{aligned}
\end{equation}

This system of linear equations involves \( n^2 \) variables with \( 3(n - 1) \) linear constraints, implying that infinitely many solutions exist in principle. However, the additional positivity constraint \( R_{ij} \geq 0 \) prevents us from directly using the linear algebra solutions.

To address this, we introduce the substitution
\begin{equation} \label{equ:9}
    R_{ij} = \frac{1}{n} + \epsilon_{ij},
\end{equation}
with \( \epsilon_{ij} \in \left[ -\frac{1}{n}, \frac{n-1}{n} \right] \), ensuring that the normalization and positivity conditions are preserved. Substituting into Eq.~\eqref{equ:8} yields the transformed system of equations
\begin{equation} \label{equ:10}
\begin{aligned}
    \sum_{j=0}^{n-1} \lambda_j \left[ \epsilon_{ij} - \left( (1 - t_i) \epsilon_{0j} + t_i \epsilon_{(n-1)j} \right) \right] = 0, \\ 
    \sum_j \epsilon_{ij} = 
    \sum_i \epsilon_{ij} = 0, \quad 
    \epsilon_{ij} \in \left[ -\frac{1}{n}, \frac{n-1}{n} \right].
\end{aligned}
\end{equation}

This is a homogeneous system of linear equations. Since homogeneous system of linear equations are scale-invariant, any solution \( \epsilon_{ij} \) can be rescaled by a constant factor without violating the equations. The interval \( \left[ -\frac{1}{n}, \frac{n-1}{n} \right] \) allows for both positive and negative values, so we can always scale a linear algebra solution such that it lies entirely within the allowed range and thus ensuring that the corresponding convex weights \( R_{ij} \) remain in the interval $[0,1]$.

\subsubsection{Linear programming problem}
In quantum sensing, the effective spectral range $\Delta^{(\text{eff})}=\lambda^{\text{(eff)}}_{\text{max}}- \lambda^{\text{(eff)}}_{\text{min}}$ directly determines the sensitivity to an external signal. Therefore, given a set of target ratios \( t_i \), our goal is to find a configuration that maximizes this spectral range. This can be naturally formulated as an optimization problem.

To proceed, we reshape the matrix \( \epsilon_{ij} \) into a vector
\[
\boldsymbol{v} = \begin{bmatrix} \epsilon_{00}, \epsilon_{01}, \dots, \epsilon_{0(n-1)}, \epsilon_{10}, \dots, \epsilon_{(n-1)(n-1)} \end{bmatrix}^T,
\]
and express the linear system of equations from Eq.~\eqref{equ:10} as a matrix equation \( A_t \boldsymbol{v} = 0 \). This enforces that \( \boldsymbol{v} \in \ker(A_t) \), subject to element-wise bounds
\[
-\frac{1}{n} \leq \boldsymbol{v}_i \leq \frac{n-1}{n}.
\]

Let \( \boldsymbol{f} \) be the vector encoding the cost function related to the spectral range $\boldsymbol{f}^T\boldsymbol{v}=\Delta^{(\text{eff})}$, explicitly 
\[
\boldsymbol{f} = \begin{bmatrix} -\lambda_0, -\lambda_1, \dots, -\lambda_{n-1}, 0, \dots, \lambda_0, \lambda_1, \dots, \lambda_{n-1} \end{bmatrix},
\]
then the optimization task becomes
\begin{align} \label{equ:11}
    &\textbf{maximize} \quad \boldsymbol{f}^T \boldsymbol{v}, \\
    &\text{subject to} \quad A_t \boldsymbol{v} = 0, \\
    &\quad\quad\quad\quad\quad -\frac{1}{n} \leq \boldsymbol{v}_i \leq \frac{n-1}{n}. \nonumber
\end{align}

To cast this as a standard linear programming (LP) problem, we switch back to the convex weights variables
\[
\tilde{\boldsymbol{v}} = \begin{bmatrix} R_{00}, R_{01}, \dots, R_{(n-1)(n-1)} \end{bmatrix}^T,
\]
and define
\[
\tilde{A} = \begin{bmatrix} A_t \\ -A_t \end{bmatrix}, \quad \tilde{\boldsymbol{b}} = \begin{bmatrix} \boldsymbol{b} \\ \boldsymbol{b} \end{bmatrix}, \quad \boldsymbol{b} = \begin{bmatrix} \boldsymbol{0}_{n-2}, \boldsymbol{1}_{2n} \end{bmatrix}^T.
\]

The problem can now be written in standard LP form
\begin{align} \label{equ:12}
    &\textbf{maximize} \quad \boldsymbol{f}^T \tilde{\boldsymbol{v}}, \\
    &\text{subject to} \quad \tilde{A} \tilde{\boldsymbol{v}} \leq \tilde{\boldsymbol{b}}, \\
    &\quad\quad\quad\quad\quad\quad\;\; \tilde{\boldsymbol{v}} \geq 0. \nonumber
\end{align}
LP problems can be solved efficiently using standard techniques, such as the simplex algorithm or interior-point methods \cite{borgwardt2012simplex,khachiyan1979polynomial}.
\subsubsection{Spectral reduction }
To determine the maximum cost of the optimization, i.e. the maximal reduction of spectral range, we can equivalently find the worst-case target vector and then calculate its corresponding spectral range.  
The worst-case target vector for any initial spectrum is one that is concentrated entirely at the edges, i.e.,
\[
\boldsymbol{t} \in \{0,1\}^{\otimes n},
\]
as such a configuration requires maximal interaction with energy levels that contribute to the reduction of the spectral range, i.e. the maximal and minimal energy levels.

Given a target vector of the form
\[
\boldsymbol{t} = (\boldsymbol{0}_m, \boldsymbol{1}_{n-m}),
\]
where $\boldsymbol{x}_l$ is a vector of length $l$ with value $x$ in each entry,
the resulting effective spectral range is
\begin{equation} \label{equ:13}
\Delta_m = \frac{1}{n - m} \sum_{i = m}^{n-1} \lambda_i - \frac{1}{m} \sum_{i = 0}^{m-1} \lambda_i,
\end{equation}
for eigenvalues ordered such that $\lambda_0 \leq \lambda_1 \leq \cdots \leq \lambda_{n-1}$.
The minimum achievable spectral range across all such edge-concentrated targets is then given by
\begin{equation} \label{equ:14}
\Delta_{\text{min}} = \min_m \Delta_m.
\end{equation}

The overall worst-case scenario, across all initial and target vectors, occurs when starting with the initial spectrum
\[
\boldsymbol{\lambda} = (\boldsymbol{0}_{n-1}, \lambda),
\]
and targeting
\[
\boldsymbol{t} = (0, \boldsymbol{1}_{n-1}).
\]

This yields an effective spectral range of
\begin{equation} \label{equ:15}
\gamma = \frac{\lambda}{n - 1},
\end{equation}
which shows that the spectrum has been compressed by a factor of \( n - 1 \). Hence, the maximal possible reduction in spectral range due to control is exactly by this factor. 
The analysis above demonstrated the worst-case performance. A numerical study of the average-case performance is presented in Appendix~\ref{App:spec_red_num_inv}.

\subsubsection{Implementation of Switching} \label{subsubsec:impl_of_switch}
\begin{figure}
    \centering
    \includegraphics[width=1\linewidth]{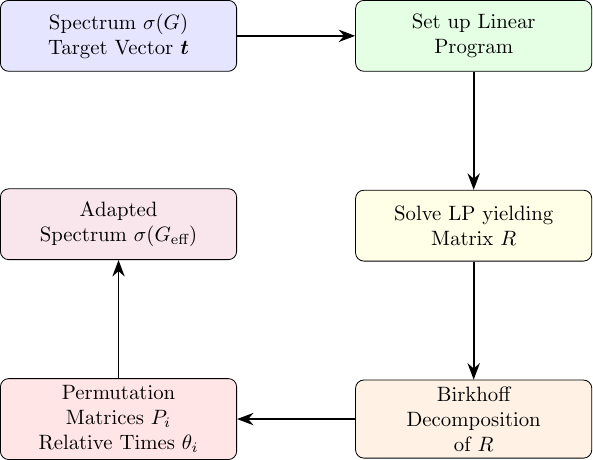}
    \caption{Illustration of the steps of the switching method to obtain from a given spectrum $\sigma(G)$ and a given target vector $\boldsymbol{t}$ the adapted spectrum $\sigma(G_\text{eff})$ and the corresponding control operations.}
    \label{fig:work_flow}
\end{figure}

Once the optimal matrix \( R_{\text{opt}} \) is found, the next step is to determine a corresponding sequence of switching operations, which implements this control operation. Since \( R_{\text{opt}} \) is a bi-stochastic matrix, it admits a so-called Birkhoff decomposition \cite{MR20547}:
\begin{equation}\label{equ:Birkhoff}
    B = \sum_i \theta_i P_i, \quad \sum_i \theta_i = 1, \quad \theta_i \geq 0,
\end{equation}
where \( P_i \) are permutation matrices, which can be decomposed into switching operations. 
The times at which the permutation operations are implemented correspond to the corresponding weight $\theta_i$ of their permutation matrix and an arbitrary order of implementation of the permutations, see Fig.~\ref{fig:circuit}(c) for an illustration. In Fig.~\ref{fig:work_flow} we summarize the steps of the switching method.

\subsubsection{Resource requirements} \label{subsubsec:resource}
Since the number of control operations is typically an important resource, we analyze how many elementary switching operations are required to realize the desired spectral modification.

The Birkhoff decomposition, see Eq.\eqref{equ:Birkhoff}, can be computed using the Birkhoff algorithm \cite{MR20547}, which terminates in at most \( \mathcal{O}(n^2) \) steps and returns no more than \( n^2 - 2n + 2 \) permutations with associated weights \cite{Johnson_Dulmage_Mendelsohn_1960}.
The Birkhoff decomposition is not unique and finding one decomposition with the minimal number of terms is known to be NP-hard \cite{DUFOSSE2016108}. Therefore, one requires up to $\mathcal{O}(n^2)$ permutation unitary to implement the desired unitary control. 

Each permutation matrix \( P_i \) can be implemented using at most \( n - 1 \) two-level permutations (i.e., switching operations). This is because, for any given permutation, we can successively move each energy level to its target position using a sequence of two-level permutations. After positioning the first \( n - 2 \) levels correctly, the remaining two levels will either already be in the correct order or can be swapped with one final operation to complete the permutation. Therefore, the full implementation of \( R_{\text{opt}} \) requires no more than $n^3 - 3n^2 + 4n - 2$ switching operations in total. Hence we require up to $\mathcal{O}(n^3)$ switching operations to implement a spectral adaptation. 

The number of required switching operations can be reduced to $\mathcal{O}(n)$ to achieve a desired target spectrum, however, this reduction comes at the cost of a potentially greater loss in spectral range, as well as a super-exponential computational cost for determining the switching sequence. The exact construction is provided in Appendix~\ref{App:min_number_of_switches}.

\subsubsection{Qubit compilation} \label{subsubsec:qubit}
If one wants to implement the permutation operations on a qubit-based quantum processor, one also needs to consider how to implement these operations efficiently. 
Note that implementing an arbitrary permutation operation on a qubit-based quantum processor is equivalent to an arbitrary reversible classical circuit. This can be readily seen via noting the effect of an classical reversible circuit is given by a function
\begin{equation}
    f:\{0,1\}^{\otimes k} \rightarrow \{0,1\}^{\otimes k},
\end{equation}
which has an inverse function $f^{-1}$. In particular this implies that $f$ is faithful so two distinct elements are mapped onto two distinct elements, and $f$ is onto so all elements $\{0,1\}^{\otimes k}$ are in the range of $f$. Therefore, the elements of the domain are simply permuted under the map $f$, and hence can be described by a permutation. On the other hand, each permutation is invertible (by its transpose) therefore it is also an invertible operation. Therefore, the question how to implement an arbitrary permutation operation is equivalent to the question of how to implement an arbitrary invertible classical circuit. In \cite{ZAKABLUKOV2017132} it was shown that for an $m$-qubit system one requires in the worst case $\mathcal{O}(2^m)$  two-qubit gates (and one-qubit gates) to implement an arbitrary reversible classical circuit. Having $n=2^m$ levels, we require $\mathcal{O}(n)$ two-qubit (and one-qubit gates) gates to implement an arbitrary permutation matrix on a qubit-based quantum processor. 

\subsection{General unitary control} \label{subsec.general_unitary_control}
In this section, we examine the effect of allowing arbitrary unitary gates to modify the spectrum of a Hamiltonian \( H \) and demonstrate that the spectral engineering capabilities reduces to the ones given by the switching method. Our goal is to adjust only the eigenvalues of \( H \) while preserving its eigenbasis. This constraint implies that the resulting effective Hamiltonian \( \tilde{H} \) must remain diagonal in the eigenbasis of \( H \).

Let the original Hamiltonian be \( H = \text{diag}(\lambda_j) \). The effective Hamiltonian under unitary control can be expressed as
\begin{equation} \label{equ:adapt}
    \tilde{H} = \frac{1}{\Delta t} \sum_i \delta_i \, U_i H U_i^\dagger,
\end{equation}
where each \( U_i \) is a unitary operation, and the infinitesimal coefficients \( \delta_i  \) denote the relative durations of the applied unitary operations, satisfying \( \sum_i \delta_i = \Delta t \), as illustrated in Fig.~\ref{fig:circuit}(b).
This result can be understood by examining the evolution operator over a time interval \( \Delta t \). We begin by splitting the evolution into weighted segments
\begin{equation}
    e^{-iH \Delta t} = e^{-i \sum_j \delta_j H } = \prod_j e^{-i \delta_j H }.
\end{equation}
Expanding each exponential using the first-order Taylor expansion yields
\begin{equation}
    e^{-iH \Delta t} = \prod_j \left( \mathbb{1} - i \delta_j H  + \mathcal{O}(\delta_j^2) \right).
\end{equation}

Now, suppose we insert unitary operations before and after each infinitesimal time step. Note this is equivalent to the insertion of arbitrary unitary operations $U_i$ before the infinitesimal time evolution and insertion of the hermitian conjugate $U_i^\dag$ after the infinitesimal evolution, since one can decompose the next arbitrary unitary $U_{i+1}=U_i^\dag \tilde{U}_{i+1}.$ This procedure yields the claimed form up to a final unitary before the evolution. This unitary can be absorbed into the probe state we are acting on via redefinition of the coefficients. The evolution becomes:
\begin{equation}
\begin{aligned}
    \prod_j U_j e^{-i \delta_j H } U_j^\dag 
    &= \prod_j \left( \mathbb{1} - i \delta_j U_j H U_j^\dag  + \mathcal{O}(\delta_j^2) \right) \\
    &=\mathbb{1} - i  \sum_j \delta_j U_j H U_j^\dag + \mathcal{O}(\delta_j^2) \\
    &= e^{-i \sum_j \delta_j U_j H U_j^\dag}.
\end{aligned}
\end{equation}

Thus, the combined effect of alternating unitary operations with infinitesimal evolutions under \( H \) leads to an effective Hamiltonian
\[
\tilde{H} = \frac{1}{\Delta t} \sum_j \delta_j U_j H U_j^\dag,
\]
which is consistent with the expression used in Eq.~\eqref{equ:adapt}.

To ensure \( \tilde{H} \) remains diagonal, only the diagonal elements of each term contribute. The diagonal entries of \( \tilde{H} \) are then
\begin{equation} \label{equ:tildelambd}
\begin{aligned}
    \tilde{\lambda}_j &= \frac{1}{\Delta t} \sum_i \delta_i (U_i H U_i^\dagger)_{jj} = \frac{1}{\Delta t} \sum_i \delta_i \sum_m |(U_i)_{jm}|^2 \lambda_m \\
    &\equiv \sum_m p_{jm} \lambda_m,
    \end{aligned}
\end{equation}
where \( p_{jm} = \frac{1}{\Delta t} \sum_i \delta_i |(U_i)_{jm}|^2 \) are the resulting weights.

We now show that \( \tilde{\lambda}_j \) is a convex combination of the original eigenvalues \( \lambda_m \). This follows from the summation condition on the infinitesimal coefficients $\delta_i$ and the fact that $U_i$ are unitary matrices
\begin{equation}
    \sum_m p_{jm} = \frac{1}{\Delta t} \sum_i \delta_i \sum_m |(U_i)_{jm}|^2 = \frac{1}{\Delta t} \sum_i \delta_i = 1,
\end{equation}
since the rows and columns of unitary matrices form orthonormal bases. Similarly holds
\begin{equation}
    \sum_j p_{jm} = \frac{1}{\Delta t} \sum_i \delta_i \sum_j |(U_i)_{jm}|^2 = \frac{1}{\Delta t} \sum_i \delta_i = 1.
\end{equation}

Therefore, the adapted eigenvalues take the form
\begin{equation}
    \tilde{\lambda}_j = \sum_m p_{jm} \lambda_m, \quad  \sum_j p_{jm} = \sum_m p_{jm} = 1.
\end{equation}

We conclude that the spectral engineering capabilities under unitary control are equivalent to those achievable using the switching method.

\section{Single parameter Bayesian estimation} \label{Sec:single_parameter_bayesian_estimation}
Here, we give the necessary theoretical background for the application of the switching method to Bayesian parameter estimation. We consider different single-shot single-parameter sensing scenarios and discuss the theoretical background.
\subsection{Frequency single-shot Bayesian estimation} \label{subsec:freq est}

In quantum metrology, one of the central tasks is to estimate an unknown physical parameter \( \omega \) that is encoded in the evolution of a quantum system. This parameter could represent a frequency, a magnetic field strength, or another physical quantity, depending on the sensing task. The setting considered here is that of single-shot estimation \cite{PhysRevA.82.053804,PhysRevLett.82.2207,berry2000optimal,PhysRevA.79.022314}, where only one round of state preparation, evolution, and measurement is performed, in contrast to repeated measurements \cite{e20090628,PRXQuantum.2.020303,vasilyev2024optimalmultiparametermetrologyquantum} where multiple rounds of state preparations and measurements are considered.

The goal in this single-shot scenario is to construct an estimation protocol comprising a quantum probe state, a measurement, and an estimator function that minimizes the uncertainty in the estimated value of $\omega$ based on prior information and a single measurement outcome. Since real-world quantum sensors often operate under constraints that limit repeated measurements, such as resource or time limitations, developing strategies optimized for the single-shot regime is practically relevant. 

To formalize the estimation task, we begin with a prior probability distribution \( p_\omega \), which encodes our initial knowledge or uncertainty about the true value of \( \omega \). The quality of an estimation strategy is quantified by the Bayesian mean squared error (BMSE) \cite{RevModPhys.83.943,van2004detection,Helstrom1969,holevo2011probabilistic}
\begin{equation} \label{equ:BMSE}
    \Delta^2 \tilde{\omega} = \int \mathrm{d}\omega\, \mathrm{d}x\, p_\omega\, \mathrm{tr}(\rho_\omega \Pi_x)\, (\omega - \tilde{\omega}_x)^2,
\end{equation}
where \( \rho_\omega \) is the evolved quantum state, \( \Pi_x \) denotes a POVM element corresponding to measurement outcome \( x \), and \( \tilde{\omega}_x \) is the estimator function assigned to outcome \( x \). Eq.~\ref{equ:BMSE} highlights the importance of the system dynamics (encoded in \( \rho_\omega \)), the measurement strategy, and the prior information for the accuracy of an estimation scheme. 

While the BMSE is the standard Bayesian cost function, other loss functions can be considered, potentially leading to different optimal protocols.

We assume the parameter is encoded through unitary evolution under a Hermitian generator \( G \), such that
\begin{equation}
    \rho_\omega = e^{-i \omega t G} \rho\, e^{i \omega t G}, \quad G = \sum_{j=0}^{n-1} \lambda_j\, \ket{j}\bra{j}.
\end{equation}

To identify the optimal measurement, we define two effective operators that encode the average state and the first moment of the parameter
\begin{equation} \label{equ:eff_states}
    \Gamma = \int \mathrm{d}\omega\, p_\omega\, \rho_\omega, \quad 
    \eta = \int \mathrm{d}\omega\, p_\omega\, \omega\, \rho_\omega.
\end{equation}

It was shown that there exist a projective measurement which minimizes the BSME, which is characterized by an operator \( L \) that satisfies the symmetric Sylvester equation \cite{1054643}
\begin{equation} \label{eq:sylvester}
    \{L, \Gamma\} = 2\eta,
\end{equation}
where \( \{ \cdot, \cdot \} \) denotes the anticommutator.

The minimal BMSE attainable with the optimal measurement and estimator is given by
\begin{equation}
    \Delta^2 \tilde{\omega} = \Delta^2 \omega - \mathrm{tr}(\Gamma L^2),
\end{equation}
where \( \Delta^2 \omega \) is the prior variance. We consider scenarios where the prior distribution is Gaussian
\begin{equation} \label{equ:gaussian_prior}
p_\omega \propto \exp\left(-\frac{\omega^2}{2 \Delta^2 \omega} \right),
\end{equation}
which is both analytically convenient and justified in many applications. 
% In particular due to the Bernstein-von Mises theorem  \cite{DasGupta2008,10.1214/12-EJS675,e20090628} in the many repetition scenario, i.e. the number of measurements $K \gg 1$, the posterior distribution converges to a normal distribution. Therefore, for the $K+1$-th measurement we are guaranteed to have a Gaussian prior distribution.

Following the framework of~\cite{macieszczak2014bayesian}, given a Gaussian prior the BMSE simplifies
\begin{equation}\label{eq:var_to_qfi}
    \Delta^2 \tilde{\omega} = \Delta^2 \omega \left[ 1 - \Delta^2 \omega\, F(\Gamma,tG) \right],
\end{equation}
where \( F(\Gamma,tG) \) is the quantum Fisher information (QFI) for an effective state \( \Gamma \)
\begin{equation} \label{equ:QFI}
    F(\Gamma,tG)=2t^2\sum_{ij} \frac{ \left|\bra{i} G\ket{j}\right|^2(\lambda_i-\lambda_j)^2}{(\lambda_i+\lambda_j)},
\end{equation}
where \( \lambda_i \) and \( \ket{i} \) are eigenvalues and eigenvectors of \( \Gamma \).

Eq.~\eqref{eq:var_to_qfi} shows that minimizing the BMSE is equivalent to maximizing the QFI for an effective state \( \Gamma \).  However, direct optimization of \( F(\Gamma,tG) \) over the full state space is computationally intensive.

To address this, we adopt an iterative scheme:
\begin{enumerate}
    \item Initialize a random pure state \( \rho_0 \).
    \item Compute the optimal measurement operator \( L_0 \).
    \item Construct the operator
    \[
    T_0 = \int \mathrm{d}\omega\, p_\omega\, e^{i t H} \left( L_0^2 - 2 \omega L_0 \right) e^{-i t H}.
    \]
    \item Set \( \rho_1 \) as the pure state corresponding to the eigenvector with the lowest eigenvalue of \( T_0 \).
    \item Repeat until convergence.
\end{enumerate}
The convergence of this algorithm was analyzed in~\cite{macieszczak2013quantumfisherinformationvariational}. 

To speed up the computation, \( \Gamma \) and \( \eta \) can be calculated analytically
\begin{equation} \label{equ:analytical_expr_spread_out_state}
\begin{aligned}
    \Gamma &= \sum_{i,j} c_i c_j^*\, \exp\left( -\frac{t^2 \Lambda_{ij}^2 \Delta^2 \omega}{2} \right) \ket{i}\bra{j}, \\
    \eta &= - i t \Delta^2 \omega \sum_{i,j} c_i c_j^* \Lambda_{ij}\, \exp\left( - \frac{t^2 \Lambda_{ij}^2 \Delta^2 \omega}{2} \right) \ket{i}\bra{j},
\end{aligned}
\end{equation}
with \( \Lambda_{ij} = \lambda_i - \lambda_j \). The Sylvester equation can be efficiently solved e.g. using the Bartels–Stewart algorithm~\cite{10.1145/361573.361582}.

Furthermore, the operators \( T_i \) used in the iteration step can be computed analytically in a similar fashion to \( \Gamma \) and \( \eta \), by identifying the term involving \( L_i^2 \) with the structure of \( \Gamma \), and the term involving \( L_i \) with that of \( \eta \).

Importantly, the QFI is a function only of \( \Gamma \), the generator \( G \), and the evolution time \( t \). In fact, by defining the dimensionless parameter \( \tau = t \Delta \omega \), we observe that
\begin{equation}
     \begin{aligned}
     F(\Gamma,tG)&=(\Delta^{ 2} \omega)^{-1} 2\tau^2\sum_{ij} \frac{ \left|\bra{i} G\ket{j}\right|^2(\lambda_i-\lambda_j)^2}{(\lambda_i+\lambda_j)} \\
     &= (\Delta^{ 2} \omega)^{-1} F(\Gamma,\tau G),
     \end{aligned}
\end{equation}
demonstrating that the QFI factorizes into a component dependent only on \( \tau \) and another proportional to the inverse prior variance. Hence Eq.~\eqref{eq:var_to_qfi} can be written as
\begin{equation}
    \Delta^2 \tilde{\omega} = \Delta^2 \omega \left[ 1 -  F(\Gamma,\tau G) \right],
\end{equation}
which implies that the qualitative behavior of the estimation strategy is determined solely by the dimensionless parameter \( \tau \), independent of the prior width. Consequently, without loss of generality, we can fix \( \Delta^2 \omega = 1 \), simplifying the analysis while preserving generality.

\subsection{Phase single-shot Bayesian estimation}

In many quantum sensing scenarios, the parameter of interest does not correspond directly to a frequency, but rather to a phase \(\phi\), which is typically accumulated over some known evolution time \(t\). Estimating this phase with high precision is central to tasks such as interferometry, clock synchronization, and rotation sensing \cite{PhysRevA.80.013825,PhysRevA.72.042301,PhysRevX.14.011033,Vidrighin2014}.

To explore the application of our method in Bayesian phase estimation, we must slightly modify the estimation framework. The discussion that follows is based on the approach introduced in~\citep{demkowicz2011optimal}, adapted here to accommodate non-linear spectra.

Assume the parameter \( \phi = \omega t \) is imprinted via a Hamiltonian of the form
\[
H = \omega \sum_j \lambda_j \ket{j} \bra{j},
\]
with the associated unitary evolution
\[
U_\phi \ket{j} = e^{-i \lambda_j \phi} \ket{j}.
\]
Given a frequency prior \(p_\omega\), the corresponding prior for the phase is given by
\begin{equation}
    p_\phi(\phi)=\sum_{k \in \mathbb{Z}} p_\omega \left(\frac{\phi+2 \pi k}{t} \right),
\end{equation}
where in the following we write \( p = p_\phi \). 
For a general quantum state \( \rho \), the evolved state is \( \rho_\phi = U_\phi \rho U_\phi^\dag \). The goal is to estimate the true value of \( \phi \) using a POVM \( \{ \Pi_{\tilde{\phi}} \} \), where \( \tilde{\phi} \) denotes the measurement outcome.

The conditional probability of measuring \( \tilde{\phi} \) given the true phase \( \phi \) is given by Born’s rule
\[
p(\tilde{\phi} | \phi) = \mathrm{tr}(\rho_\phi \Pi_{\tilde{\phi}}).
\]

Given a prior distribution \( p(\phi) \), the average cost associated with a given estimation strategy is
\begin{equation}
\bar{C} := \int_{-\pi}^{\pi} d\phi\, d\tilde{\phi}\; p(\phi)\, \mathrm{tr}(\rho_\phi \Pi_{\tilde{\phi}})\, C(\phi, \tilde{\phi}),
\end{equation}
where \( C(\phi, \tilde{\phi}) \) denotes the cost of estimating \( \tilde{\phi} \) when the true value is \( \phi \).

While the quadratic cost \( C(\phi, \tilde{\phi}) = (\phi - \tilde{\phi})^2 \) (cf. BMSE Eq.~\eqref{equ:BMSE}) is appropriate for narrow distributions, it fails to account for the periodicity of the phase. Therefore, we instead adopt the periodic cost function
\[
C(\phi, \tilde{\phi}) = 4 \sin^2\left(\frac{\phi - \tilde{\phi}}{2}\right),
\]
which is the simplest cost function (smallest number of Fourier coefficients) which approximates the variance for narrow distributions \cite{berry2000optimal} and we denote its average with $\delta^2 \tilde{\phi}$ and interpret it as a measure for the estimation error.

This cost function admits a useful operator representation
\[
C(\phi, \tilde{\phi}) = 4 \left(1 - \mathrm{tr}\left[\ket{\phi}\bra{\phi} \ket{\tilde{\phi}}\bra{\tilde{\phi}} \right] \right),
\]
where
\[
\ket{\phi} := \frac{1}{\sqrt{2}} \left( \ket{0} + e^{-i\phi} \ket{1} \right).
\]

To proceed, we define the operators
\begin{align}
R &= \int_{-\pi}^{\pi} d\phi\; p(\phi) \ket{\phi}\bra{\phi} \otimes \rho_\phi, \\
M &= \int_{-\pi}^{\pi} d\tilde{\phi}\; \ket{\tilde{\phi}}\bra{\tilde{\phi}} \otimes \Pi_{\tilde{\phi}},
\end{align}
so that the average cost takes the form
\begin{equation} \label{eq:min_av_cost}
\delta^2 \tilde{\phi} = 4(1 - F), \quad F = \mathrm{tr}(R M).
\end{equation}

Next, we perform a partial trace over the auxiliary space (associated with the \(\ket{\phi}\) register), defining:
\[
R_i^j := \bra{i}_A R \ket{j}_A, \quad M_i^j := \bra{i}_A M \ket{j}_A.
\]

Assuming the prior \( p(\phi) \) has the following Fourier expansion
\[
p(\phi) = \frac{1}{2\pi} \sum_{k \in \mathbb{Z}} p_k e^{i k \phi},
\]
and using the fact that the POVM must satisfy completeness (\( \mathrm{tr}_A M = \mathbb{1} \)), we obtain
\[
M_0^0 + M_1^1 = \mathbb{1}, \quad R_0^0 = R_1^1, \quad \mathrm{tr}(R_i^i) = \frac{1}{2}.
\]

Using these identities, we can express \( F \), see Eq.~\eqref{eq:min_av_cost}, as
\[
F = \frac{1}{2} + 2\, \mathrm{Re}\left[\mathrm{tr}(R_0^1 M_1^0)\right],
\]
and the average cost can be written as
\begin{equation}
\delta^2 \tilde{\phi} = 4\left( \frac{1}{2} - \| R_0^1 \|_1 \right),
\end{equation}
where \( \| \cdot \|_1 \) is the trace norm. Thus, the optimal estimation strategy is the one that maximizes the trace norm of \( R_0^1 \).

The optimal measurement achieving this minimum cost has the form
\begin{equation}
M = \sum_k \ket{\phi_k} \bra{\phi_k} \otimes \ket{\Psi_k} \bra{\Psi_k},
\end{equation}
where \( e^{-i \phi_k} \) and \( \ket{\Psi_k} \) are the eigenvalues and eigenvectors of the unitary operator
\[
U = V_R U_R^\dag,
\]
obtained from the singular value decomposition
\[
R_0^1 = U_R \Lambda_R V_R^\dag.
\]

Finally, the explicit matrix elements of \( R_0^1 \) are given by:
\begin{equation} \label{equ: r_matrix}
(R_0^1)_j^i = \frac{1}{2} \rho_{ij} \sum_{k \in \mathbb{Z}} p_k\, \text{sinc} \left[ \pi (k + 1 - (\lambda_i - \lambda_j)) \right].
\end{equation}

This expression reflects how the optimal estimation strategy depends jointly on the spectral structure of the Hamiltonian (through \( \lambda_i - \lambda_j \)), the properties of the probe state (through \( \rho_{ij} \)), and the Fourier components of the prior.

\section{Application of Method}\label{Sec:Application of method}
Here we consider different physical set-ups and estimations tasks to illustrate the power of the method to enhance sensing performance.
\subsection{Degeneracy lifting } \label{subsec:deg_lifting}
\begin{figure}
    \centering
    \includegraphics[width=1\linewidth]{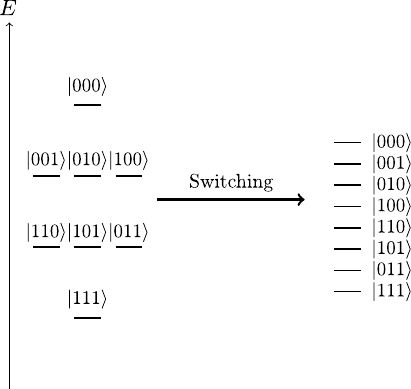}
    \caption{Sketch of a possible switching-method lifting of a three-qubit degeneracy.
Left: the degenerate spectrum defined by Eq.~\eqref{eq:deg_ham}.
Right: an example non-degenerate spectrum obtained after applying the switching protocol. }
    \label{fig:ill_deg_lifting}
\end{figure}
Here we demonstrate that lifting spectral degeneracies, see as an example Fig.~\ref{fig:ill_deg_lifting},  using the switching method can enhance quantum sensing performance. To illustrate this, we consider the Hamiltonian of $m$ qubit sensors
\begin{equation}\label{eq:deg_ham}
    H = \omega \sum_{i=1}^{m} Z_i,
\end{equation}
where \( Z_i \) are local Pauli-\( Z \) operators, corresponding to the local generator of each sensing qubit. The spectrum of this Hamiltonian consists of \( m+1 \) distinct eigenvalues
\[
\sigma\left(\sum_{i=1}^{m} Z_i\right) = \{-m, -(m-2),  \dots, m-2, m\},
\]
and hence is a degenerate spectrum, as there are in total $2^m$ eigenvalues.
We want to estimate the parameter $\omega$, with a prior distribution given by the normal distribution Eq.~\eqref{equ:gaussian_prior} with (w.l.o.g. see Sec.~\ref{subsec:freq est}) unit prior width.

\begin{figure}[ht]
    \centering
    \includegraphics[width=1\linewidth]{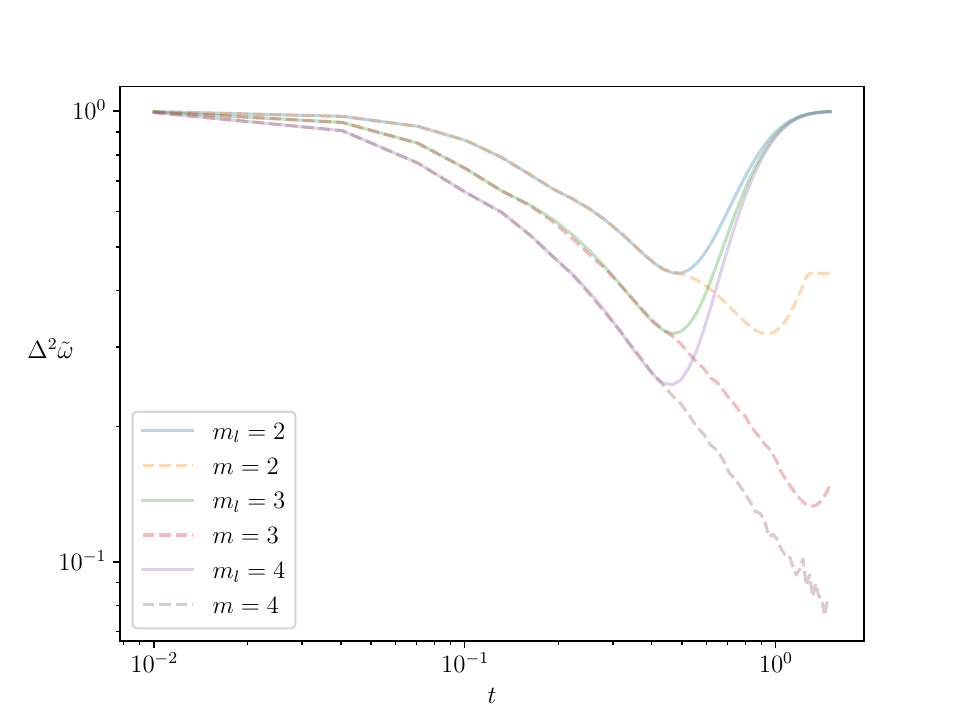}
    \caption{BMSE $\Delta^2 \tilde{\omega}$ of the optimal Bayesian estimation strategy for a Gaussian prior with unit width. Results are shown for the degenerate spectrum of the Hamiltonian in Eq.~\eqref{eq:deg_ham} (labeled \( m_l \)) and for the optimized, degeneracy-lifted spectrum obtained via the switching method (labeled \( m \)) across different system sizes $m$ ($m_l$).}
    \label{fig:degenerate}
\end{figure}

We apply the switching method to lift the spectral degeneracies and numerically optimize the target spectrum to enhance the estimation performance. The BMSE is used as a figure of merit and evaluated over time for different system sizes \( m \), as shown in Fig.~\ref{fig:degenerate}.

Initially, both the degenerate and optimized spectra yield similar performance. However, as time increases and phase wrapping effects begin to degrade the estimation accuracy in the degenerate case, the adapted spectrum continues to gain accuracy. This demonstrates that lifting degeneracies can significantly extend the useful sensing time window and hence decrease the minimal BMSE.

\subsection{Modify non-linear atomic spectrum}
\begin{figure}
    \centering
    \includegraphics[width=1\linewidth]{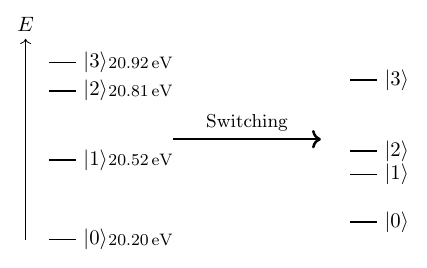}
    \caption{Sketch of a possible switching-method adaptation of the Rb III 4F-shell spectrum.
Left: the reference energies measured by Sansonetti \cite{sansonetti2006wavelengths}.
Right: an example adjusted spectrum obtained after applying the switching protocol.}
    \label{fig:ill_atom}
\end{figure}
\begin{figure}[ht]
    \centering
    \includegraphics[width=1\linewidth]{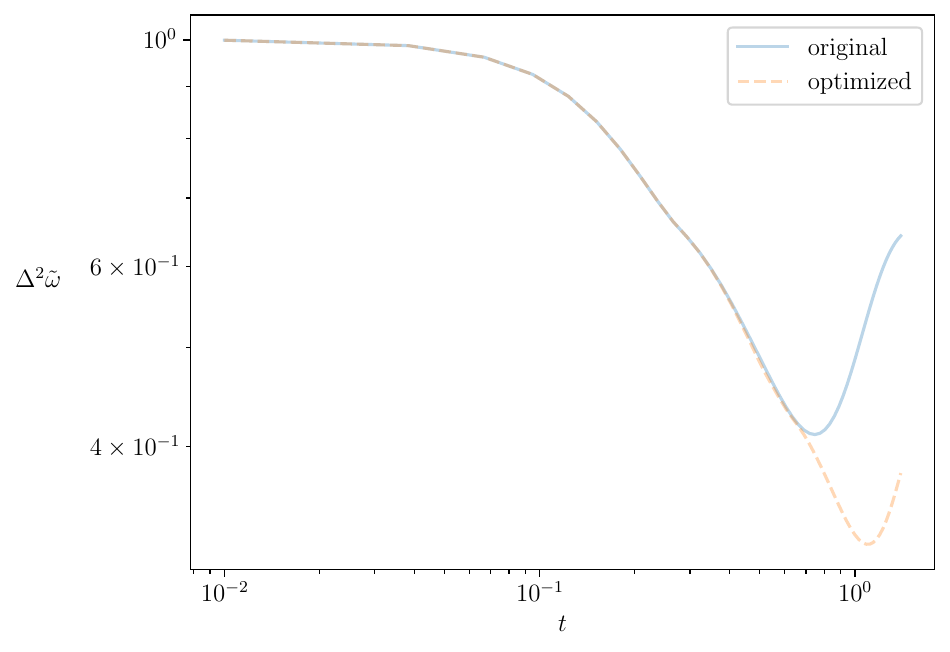}
    \caption{BMSE of the optimal Bayesian estimation strategy for a Gaussian prior with unit width. Results are shown for the original Rb~\rom{3} 4F shell spectrum and for the spectrum optimized via the switching method.}
    \label{fig:rbIII}
\end{figure}

We now investigate the impact of spectral optimization on a non-degenerate, non-linear, physically motivated spectrum. We consider the energy spectrum of the Rubidium atom, which is commonly used in many metrology applications \cite{Feng2025,Bai_2021,korth16}. Specifically, we consider the energy levels of the Rb~\rom{3} 4F~shell, as reported in~\cite{sansonetti2006wavelengths}, see Fig.~\ref{fig:ill_atom} for an illustration. We want to estimate the parameter $\omega$, with a prior distribution given by the normal distribution Eq.~\eqref{equ:gaussian_prior} with (w.l.o.g. see Sec.~\ref{subsec:freq est}) unit prior width. Fig.~\ref{fig:rbIII} compares the original Rb~III spectrum with one optimized via the switching method, where the optimization optimizes over target spectra to minimize the BMSE.

Initially, both spectra yield comparable BMSE. However, as time progresses and phase wrapping begins to degrade the performance of the original spectrum, the optimized spectrum continues to increase estimation accuracy. This highlights that even for non-degenerate spectra, targeted spectral engineering through switching can extend the useful sensing regime and hence decrease the minimal BMSE.

\subsection{Modify linear spectrum}\label{subsubsec:linear}
\begin{figure}
    \centering
    \includegraphics[scale=1.2]{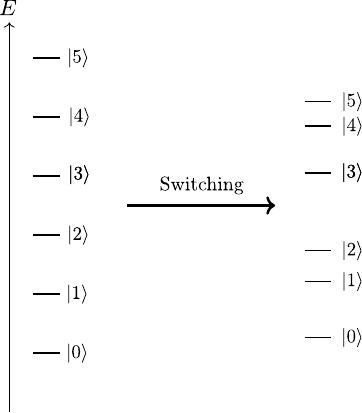}
    \caption{Sketch of a possible switching-method adaptation of the linear spectrum.
Left: the original equally spaced levels defined by Eq.~\eqref{eq:ham_linear}.
Right: an example spectrum obtained after applying the switching protocol.}
    \label{fig:ill_linear}
\end{figure}
\begin{figure}
    \centering
    \includegraphics[width=1\linewidth]{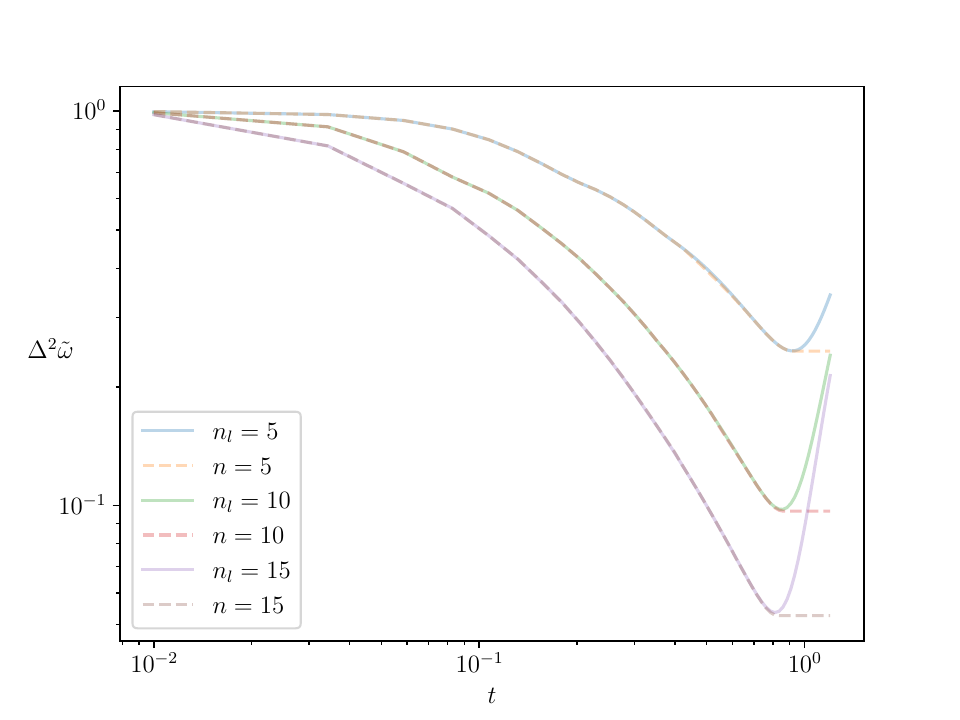}
    \includegraphics[width=1\linewidth]{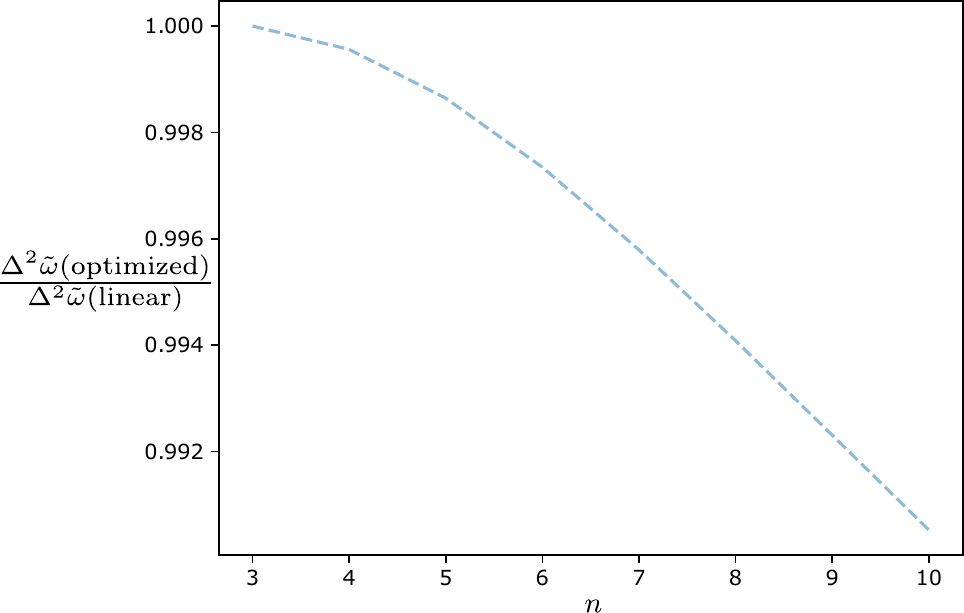}
    \caption{
    \textbf{Top}: BMSE $\Delta^2 \tilde{\omega}$ of the optimal Bayesian estimation strategy for a Gaussian prior with unit width. Results are shown for the linear spectrum defined in Eq.~\eqref{eq:ham_linear} (labelled \( n_l \)) and for the spectrum optimized via the switching method (labelled \( n \)),  as a function of the number of levels \( n \) ($n_l$).
   \textbf{Bottom}: Relative BMSE of the optimal strategies at the optimal sensing time, comparing the linear spectrum with the numerically optimized spectrum (with bounded spectral range) as a function of the number of levels \( n \).
   } 
    \label{fig:linear_spec}
\end{figure}

A widely used reference in quantum metrology is the linear spectrum, defined by the Hamiltonian
\begin{equation} \label{eq:ham_linear}
    H = \omega \sum_{i=0}^{n-1} i\, \ket{i}\bra{i}.
\end{equation}
Linear spectra appear naturally in quantum mechanical systems, such as trough the splitting of degenerate energy levels due to the Zeeman or Stark effect as well as in quantum harmonic oscillators.   
We want to estimate the parameter $\omega$, with a prior distribution given by the normal distribution Eq.~\eqref{equ:gaussian_prior} with (w.l.o.g. see Sec.~\ref{subsec:freq est}) unit prior width.
In Fig.~\ref{fig:linear_spec}~(top), we compare this standard linear spectrum with one optimized via the switching method, see Fig.~\ref{fig:ill_linear} for an illustration, where we optimize over the target spectra to minimize the BMSE.
The results show that, across all time scales, the optimized spectrum does not significantly outperform the linear spectrum in terms of BMSE. Additionally, when comparing the BMSE at their respective optimal sensing times, the improvement obtained by optimizing over all spectra constrained within the spectral range of the original spectrum is less than $1\%$ for level numbers $n \in [3,10]$, as shown in Fig.~\ref{fig:linear_spec}~(bottom).

\subsection{Adaptation of the spectrum to the prior}
\begin{figure}
    \centering
    \includegraphics[width=1\linewidth]{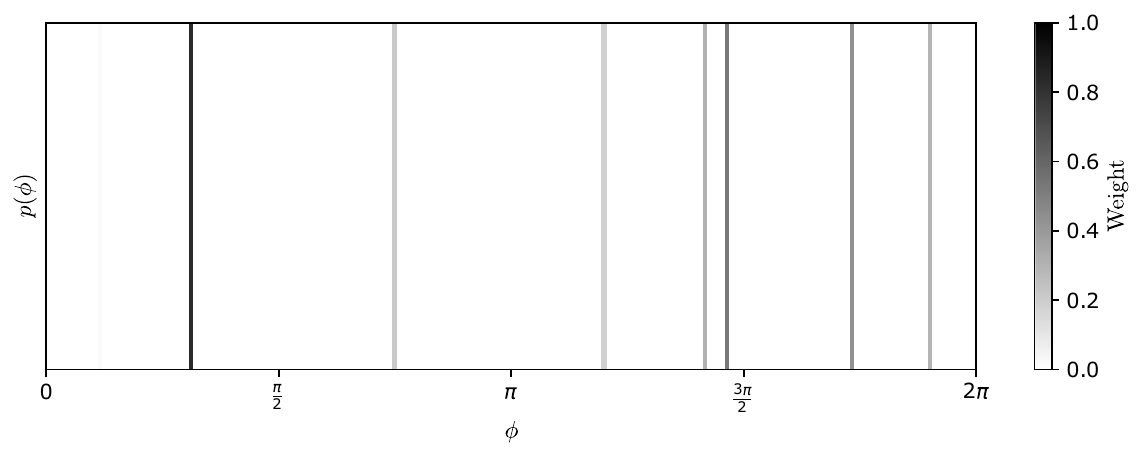}
    \caption{Illustration of a point-mass (delta-peak) prior given by Eq.~\eqref{equ:prior_delta} encoding the information of the phase $\phi$.}
    \label{fig:delta_ill}
\end{figure}
The optimal spectrum for Bayesian estimation can strongly depend on the form of the prior distribution. As shown in Sec.~\ref{subsubsec:linear}, for Gaussian priors, the standard linear spectrum performs nearly optimally, with only marginal improvements achievable through optimization. However, this is not necessarily the case for more structured priors.
To demonstrate this, we want to estimate a phase $\phi$ with a point-mass (delta-peak) prior of the form:
\begin{equation} \label{equ:prior_delta}
    p(\phi) = \sum_l w_l \delta(\phi - x_l), \quad \sum_l w_l = 1, \quad w_l \geq 0,
\end{equation}
see Fig.~\ref{fig:delta_ill} for an example.
For such a prior, the matrix elements of \( R_0^1 \) become:
\begin{equation} \label{eq:r_part_prior}
    (R_0^1)_{j}^i = \frac{1}{2} \rho_{ij} \sum_l w_l\, e^{-i x_l(\lambda_i - \lambda_j + 1)}.
\end{equation}
Minimizing the average cost \(\delta^2 \tilde{\phi} \) (cf. Eq.~\eqref{eq:min_av_cost}) is thus equivalent to maximizing the trace norm of the matrix in Eq.~\eqref{eq:r_part_prior}.

\begin{figure}
    \centering
    \includegraphics[width=1\linewidth]{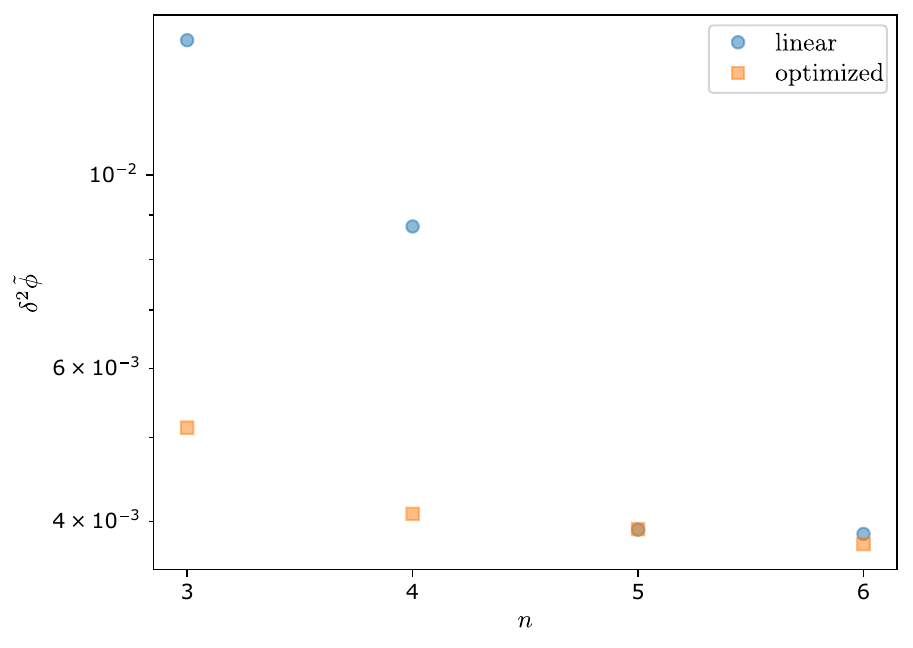}
    \caption{
    % \textbf{Top:} Point-mass prior composed of three delta peaks located at \( \phi = 0,\, 0.8,\, 2.1 \). 
    Average cost $\delta^2 \tilde{\phi}$ for prior $ p(\phi) = 0.34\delta(\phi - 2.1) + 0.15\delta(\phi + 2.5) + 0.51\delta(\phi + 2.7)$ of the optimal estimation strategy for the linear spectrum and for the numerically optimized spectrum obtained via the switching method. 
    }
    \label{fig:delta}
\end{figure}

As an example, we study a prior consisting of three weighted delta peaks given as
\begin{equation}
    p(\phi) = 0.34\delta(\phi - 2.1) + 0.15\delta(\phi + 2.5) + 0.51\delta(\phi + 2.7).
\end{equation}
We compare the performance of the switching-based spectral optimization with that of the standard linear spectrum for number of levels going from \( n = 3 \) to \( n = 6 \). As shown in Fig.~\ref{fig:delta}, the optimized spectrum consistently yields a lower average cost $\delta \tilde{\phi}^2$. In particular, for small $n$ (\( n = 3 \) and \( n = 4 \)), the improvement is substantial, demonstrating that prior-aware spectral design can significantly enhance estimation performance in the presence of structured priors.

\subsection{Flat prior} \label{Sec:flat_prior}
\begin{figure}
    \centering
    \includegraphics[width=1\linewidth]{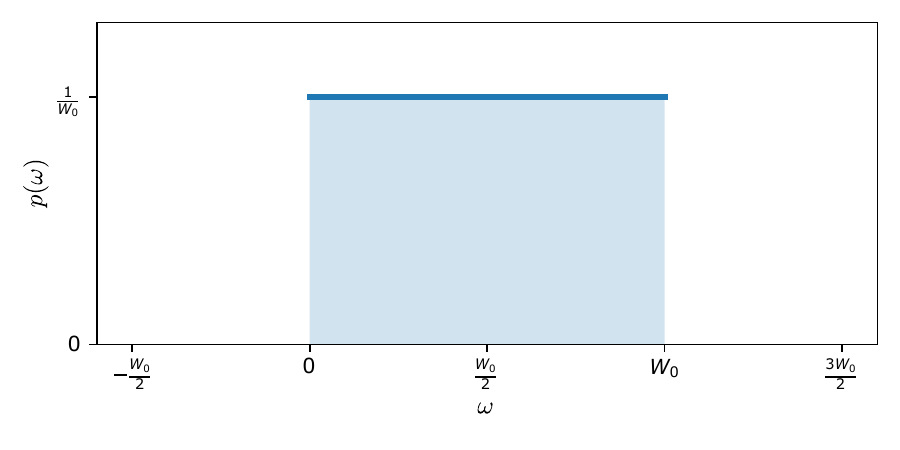}
    \caption{Illustration of the flat prior  $p(\omega)= \chi_{[0,W_0]}(\omega)/W_0$ encoding the prior knowledge of the frequency parameter $\omega$. }
    \label{fig:flat_ill}
\end{figure}
We analyze the effect of spectral spacings on frequency estimation performance for flat priors, i.e. $p(\omega)= \chi_{[0,W_0]}(\omega)/W_0$ with $\chi$ being the indicator function, see Fig.~\ref{fig:flat_ill} for an illustration. We show that the optimal spectrum for the optimal time is given by the linear spectrum.

To this aim we initially examine the scenario of phase estimation, i.e. $p(\phi)=1/(2 \pi)$,  subsequently relating the results back to frequency estimation. A central result we establish is that the linear spectrum is optimal among all integer-valued spectra.

For a linear spectrum, the matrix elements of the operator $R_0^1$ (cf. Eq.~\eqref{equ: r_matrix}) are given by
\begin{equation}
(R_0^1)_j^i = \frac{1}{2} c_i c_{i-1}^{*} \delta_{i-1,j}.
\end{equation}

For a general integer spectrum, i.e. the eigenvalues are out of $\mathbb{Z}$, the elements take the form
\begin{equation}
(R_0^1)_j^i = \frac{1}{2} c_i c_{i-1}^{*} \delta_{\Lambda_{i,j}-1,0},
\end{equation}
where $\Lambda_{i,j}$ represents the eigenvalue difference associated with states $\ket{i}$ and $\ket{j}$.

The trace norm of $R_0^1$ quantifies estimation performance. Explicitly, for the linear case,
\begin{equation}
\| R_0^1 \| = \frac{1}{2} \sum_{j} |c_j||c_{j+1}|,
\end{equation}
and in the integer case,
\begin{equation}
\| R_0^1 \| = \frac{1}{2} \sum_{j} |c_j||c_{j+1}| \delta_{\Lambda{j+1,j},1}.
\end{equation}

Thus, the trace norm in the integer scenario is always upper bounded by the linear case, with equality only when the integer spectrum is linear. Consequently, for flat priors, a linear spectrum ensures optimal phase estimation performance among integer spectra.

In \cite{wolk2020noisy}, the optimal sensing results for phase estimation with a flat prior derived in \cite{berry2000optimal} are generalized to frequency estimation using a rescaled linear spectrum. Following the approach in \cite{wolk2020noisy} we consider the linear generator for phase estimation of $m$ qubits which is given by
\begin{equation}
G_{BW} = \sum_{j=1}^{m+1} j \ket{j}\bra{j},
\end{equation}
while the corresponding rescaled generator for frequency estimation is
\begin{equation}
G = \frac{\Delta}{n-1} \sum_{j=1}^n j \ket{j}\bra{j},
\end{equation}
where $n=m+1$ denotes the number of levels.
Time evolution under $G$, described by $e^{-i \omega t G}$, maps to a phase evolution under $G_{BW}$ with an effective phase $\phi = \frac{\omega t \Delta}{n - 1}$. Assuming a uniform frequency distribution $\omega \in [0, W_0]$ and setting interrogation time
\begin{equation}
t = \frac{2\pi (n - 1)}{W_0 \Delta},
\end{equation}
yields a uniform phase distribution $\phi \in [0, 2\pi]$.

Under these conditions, the phase estimation precision for a flat prior is
\begin{equation}
\delta^2 \tilde{\phi} = \frac{\pi^2}{n^2} + \mathcal{O}(n^{-4}),
\end{equation}
translating to a frequency estimation error of
\begin{equation}
\Delta^2 \tilde{\omega} = \frac{\delta^2 \tilde{\phi}}{|\partial_\omega \phi|^2} = \frac{W_0^2}{4 n^2} + \mathcal{O}(n^{-4}).
\end{equation}

Now consider an arbitrary rescaled integer spectrum,
\begin{equation}
\tilde{G} = \frac{\Delta}{n - 1} \sum_{j=1}^{n} \lambda_j \ket{j}\bra{j},
\end{equation}
with corresponding phase generator
\begin{equation}
\tilde{G}_{BW} = \sum_{j=1}^{n} \lambda_j \ket{j}\bra{j},
\end{equation}
where $\lambda_i \in \mathbb{Z}$.
With the optimal interrogation time, the frequency estimation error satisfies
\begin{equation}
\Delta^2 \tilde{\omega} = \frac{\delta^2 \tilde{\phi}_{\text{integer}}}{|\partial_\omega \phi|^2} \geq \frac{\delta^2 \tilde{\phi}_{\text{linear}}}{|\partial_\omega \phi|^2}.
\end{equation}

As scaled integer spectra approximate any spectrum arbitrarily well, since decimal numbers are dense in the real numbers, the linear spectrum is optimal for frequency estimation given a flat prior. Thus, adjusting spectra using the switching method, from nonlinear to linear, consistently improves optimal time estimation performance, however, potentially requiring longer interrogation periods.

\subsection{Auxiliary systems}
\begin{figure}
    \centering
    \includegraphics[width=1\linewidth]{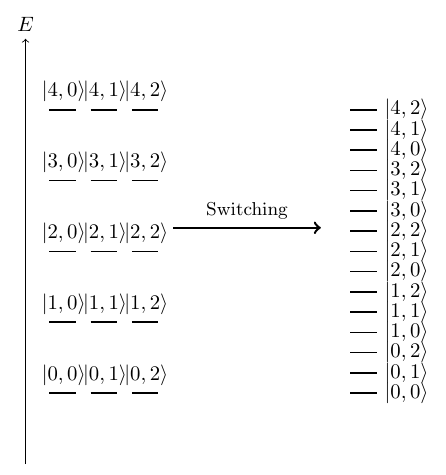}
    \caption{Sketch of switching-method adaptation of a linear spectrum endowed with an auxiliary system.
Left: the linear spectrum endowed with an auxiliary system defined by  Eq.~\eqref{equ:aux_ham}.
Right: the non-degenerate spectrum obtained after applying the switching protocol.  }
    \label{fig:ill_aux}
\end{figure}
Here, we allow for the employment of auxiliary systems, i.e. quantum subsystems available for performing operations but not directly sensing. In \cite{PhysRevLett.118.170801} it was already shown that using auxiliary systems one can increase the useful interrogation time of a single qubit system. Here we demonstrate that this effect can be reproduced using the switching method, as well as, we demonstrate that also for initial qudit systems we can enhance the useful interrogation time. We can model an auxiliary system of dimension $d_A$ via augmenting our sensing Hamiltonian $H_{\text{sens}}$ via
\begin{equation}
    H_{\text{total}}=H_{\text{sens}} \otimes \mathbb{1}_{d_A}=\omega \sum_i \lambda_i \ket{i}\bra{i} \otimes \mathbb{1}_{d_A},
\end{equation}
which introduces a $d_A$-fold degeneracy of each eigenvalue of the sensing Hamiltonian. Therefore, the total Hamiltonian can be written as 
\begin{equation}\label{equ:aux_ham}
     H_{\text{total}}=\omega \sum_i \lambda_i \sum_{j=0}^{d_A-1} \ket{i,j}\bra{i,j}.
\end{equation}

\begin{figure}
    \centering
    \includegraphics[width=1\linewidth]{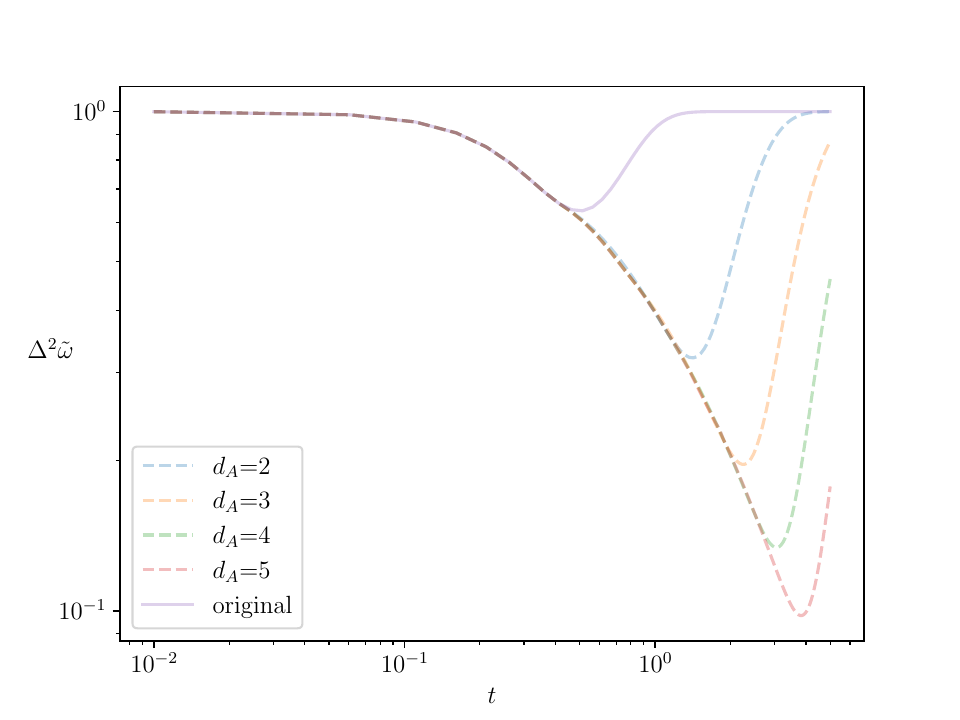}
    \includegraphics[width=1\linewidth]{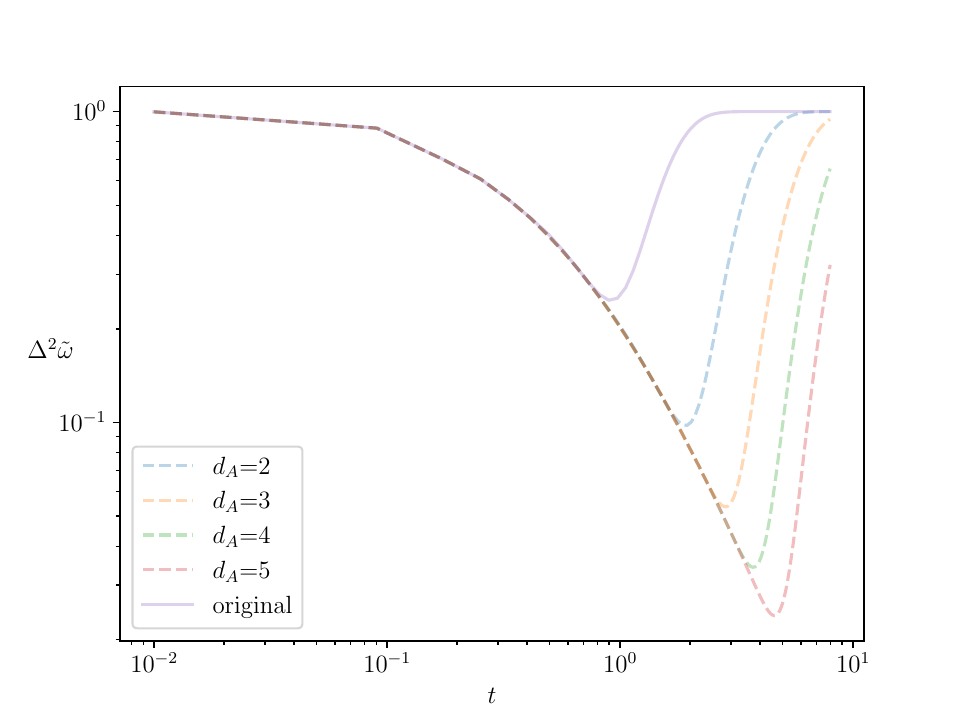}
    \caption{BMSE $\Delta^2 \tilde{\omega}$ of a sensing system (\textbf{Top}: qubit \textbf{Bottom}: five-level linear spectrum) being enhanced by auxiliary systems of dimension $d_A$, via moving to a linear spectrum with the switching method.}
    \label{fig:aux}
\end{figure}

Here we consider two examples: one being a qubit sensor enhanced by auxiliary systems and the other a 5-level linear spectrum enhanced by auxiliary systems. Note that moving from a degenerate linear system to a non-degenerate linear system with up to $d_An$ levels, we have no cost in the spectral range, see Fig.~\ref{fig:ill_aux} for an illustration.
We are considering (w.l.o.g. see Sec.~\ref{subsec:freq est}) a unit width Gaussian prior and as seen in Sec.~\ref{subsubsec:linear} the linear spectrum is almost optimal for this sensing task. Therefore, we restrict ourselves to adapting to a linear spectrum. In Fig.~\ref{fig:aux} one observes that the sensing performance of the sensing systems with incorporation of auxiliary systems is initially the same until the phase wraps start to effect the system without auxiliary systems, there the higher dimensional systems start to outperform the lower dimensional systems. Note that the drop in the estimation accuracy, due to phase wraps, could be prevented via effectively stopping the sensing via switching operations. 

\section{Conclusion} \label{Sec: Conclusion}
In this paper, we apply fast unitary control to manipulate the imprinting process of a parameter encoded in a Hamiltonian by adapting the effective spectrum of the Hamiltonian, aiming to enhance the estimation performance for this parameter. We demonstrate that spectrum manipulation using generic fast unitary control can be simplified to a finite number ($\mathcal{O}(n^3)$) of elementary switching operations. Moreover, we show that, at the possible cost of reduction of spectral range, any relative spectrum can be achieved and that the effective spectrum can be represented as a convex combination of the original eigenvalues, where the corresponding convex weights form a bi-stochastic matrix. Using this spectral engineering approach, we investigate several examples within single-parameter (phase) estimation and illustrate the effectiveness in improving estimation performance.

Here, we limited our exploration to single-parameter, single-shot estimation scenarios. Thus, a natural extension would be to investigate the method's applicability in multi-parameter estimation contexts. Additionally, since our examples have focused on single-shot estimation, exploring how the switching method can be combined with adaptive multi-shot sequences presents a promising direction.

Furthermore, our analysis considered only decoherence-free estimation cases. However, practical sensing tasks involve noise, negatively impacting sensing performance. To mitigate this, employing quantum error correction techniques \cite{PhysRevLett.112.080801,PhysRevLett.112.150801,PhysRevX.7.041009,Sekatski2017quantummetrology,Zhou2018} or decoherence-free subspaces \cite{Hamann_2024,wolk2020noisy,Hamann_2022,PhysRevResearch.2.023052,Hamann_2025} can help achieve optimal sensing outcomes. It would be valuable to explore how spectral engineering techniques could be effectively integrated with these noise-resilient strategies.

Another promising research direction involves applying the proposed method within distributed sensing frameworks, where constraints such as locality and entanglement cost become significant considerations.

Lastly, the general task of optimizing estimation strategies typically requires determining the optimal measurement, state, and estimator. With the spectral engineering method, one must additionally optimize the effective spectra. Therefore, developing efficient methods for computing these optimal effective spectra represents an area for future investigation.

\section*{Acknowledgments}
We would like to thank Denis V. Vasilyev for their valuable feedback on the initial draft of this manuscript. The authors acknowledge support from the Austrian Science Fund (FWF). This research was funded in whole or in part by the Austrian Science Fund (FWF) 10.55776/P36009, 10.55776/P36010 and 10.55776/COE1. For open access purposes, the author has applied a CC BY public copyright license to any author accepted manuscript version arising from this submission. Finanziert von der Europäischen Union.

% Bibliography
\normalem % This makes the book title in italics

\bibliographystyle{apsrev4-1}
\bibliography{ref.bib}

\clearpage

%% Appendix
\renewcommand\appendixname{Appendix}
\appendix
\onecolumngrid

\section{Spectral reduction: Numerical investigation} \label{App:spec_red_num_inv}
\begin{figure}[ht!]
    \centering
    \subfloat[]{\label{fig:average}\includegraphics[width=0.45\columnwidth]{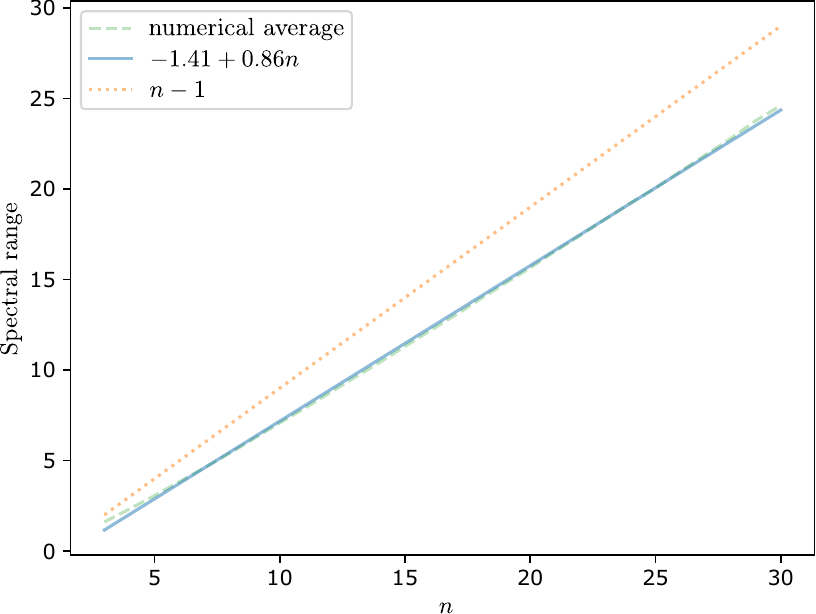}} 
    \subfloat[]{\label{fig:average_minimum}\includegraphics[width=0.45\columnwidth]{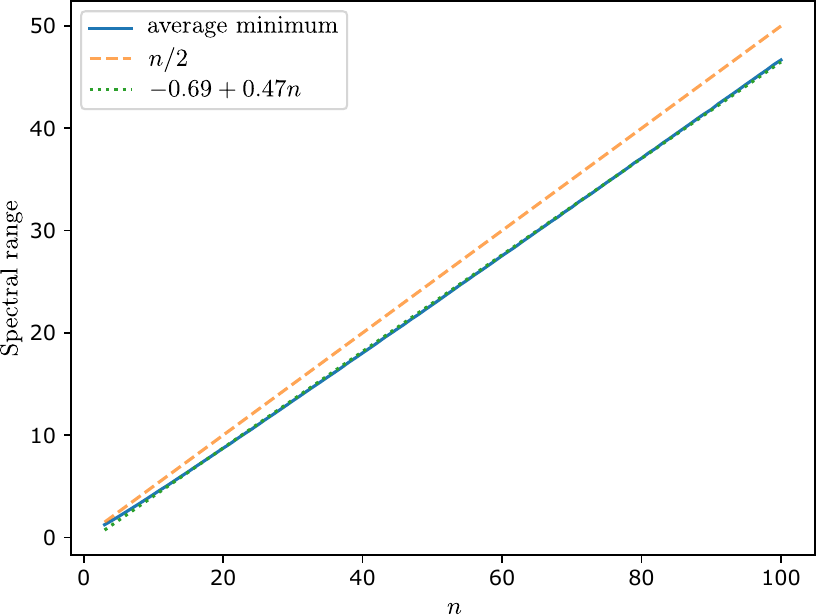}}
    \caption{ Fig.~(a) shows the average spectral range for each \( n \), computed over $10^4$ randomly sampled initial spectra (each with spectral range \( n - 1 \)) and random target vectors. Fig.~(b) displays the average minimum achievable spectral range for each \( n \), computed over $10^4$ random initial spectra with spectral range \( n - 1 \).
  }
    \label{fig:avg_spec_range}
\end{figure}

Here, we numerically investigate the spectral reduction caused by the adaptation of the spectrum. Fig.~\ref{fig:average} presents the average spectral range achieved for different values of \( n \), computed over $10^4$ randomly generated initial spectra with spectral range \( n - 1 \), each paired with a randomly sampled target vector. This provides a statistical estimate of the typical spectral performance across random configurations. A linear fit shows that the average spectral range grows almost linearly in $n$ with a slope of $0.86$. In contrast, Fig.~\ref{fig:average_minimum} shows the average minimum spectral range achievable for each \( n \), again based on $10^4$ randomly sampled initial spectra. This reflects the worst-case performance achievable through spectral engineering for random inputs. Here, a linear fit reveals that the average minimum spectral range also behaves almost linearly with a slope of $ 0.47$. Together, the figures illustrate how the achievable spectral range scales with system size, both for average-case and minimal configurations. In particular, we see that for a random task the spectral reduction is smaller than for the worst case scenario. Furthermore, we observe that the typical smallest spectral range is bigger than the global worst case given in Eq.~\eqref{equ:15}.

\section{Minimizing number of switching operations} \label{App:min_number_of_switches}
While unitary control enables broad spectral engineering, it is itself a valuable resource. Therefore, it is natural to ask how the number of required interactions can be minimized. One approach is to start with an optimal bi-stochastic matrix \( R_{\text{opt}} \) (i.e. one that maximizes spectral range), and then find its minimal Birkhoff decomposition (see Sec.~\ref{subsubsec:resource}).

Alternatively, if the goal is simply to realize a given target spectrum \( \boldsymbol{t} \) using the fewest possible interactions, we can reverse the process. Any bi-stochastic matrix \( R \) can be written as
\begin{equation}
    R= r_0 \mathbb{1} + \sum_{i=1}^k r_i P_i, \quad r_i \geq 0, \quad \sum_{i=0}^k r_i = 1,
\end{equation}
where \( P_i \) are permutation matrices. The target spectrum constraints then take the form:
\begin{equation}
    \sum_{i=0}^k r_i \left[ \sigma^i(\lambda_j) - (1 - t_j)\sigma^i(\lambda_0) - t_j \sigma^i(\lambda_{n-1}) \right] = 0, \quad \sum_i r_i = 1,
\end{equation}
where \( \sigma^0(i) = i \), and \( \sigma^i \) is the permutation associated with \( P_i \). Since there are \( n - 1 \) independent equations, at least \( k = n - 2 \) parameters are required.

To use linear algebra solutions, we introduce the substitution
\begin{equation}
    r_i = \frac{1}{n - 1} + \epsilon_i,
\end{equation}
yielding the system
\begin{equation}
    \sum_{i=0}^{n-2} \epsilon_i \left[ \sigma^i(\lambda_j) - (1 - t_j)\sigma^i(\lambda_0) - t_j \sigma^i(\lambda_{n-1}) \right] = 0, \quad \sum_i \epsilon_i = 0.
\end{equation}

Each permutation matrix \( P_i \) can be implemented using at most \( n - 1 \) pairwise (two-level) permutations. Thus, in general, the number of required switching operations is \( \mathcal{O}(n^2) \). However, we can reduce this further by constraining \( P_i \) to be products of two-level permutations
\begin{equation} \label{equ:min}
    P_1 = P_1^{(2)}, \quad P_2 = P_1^{(2)} P_2^{(2)}, \quad \dots, \quad P_{n-2} = \prod_{i=1}^{n-2} P_i^{(2)},
\end{equation}
where each \( P_i^{(2)} \) is a single two-level (flipping) permutation. This strategy reduces the switching cost to \( \mathcal{O}(n) \).

\begin{figure}
    \centering
    \includegraphics[width=0.5\linewidth]{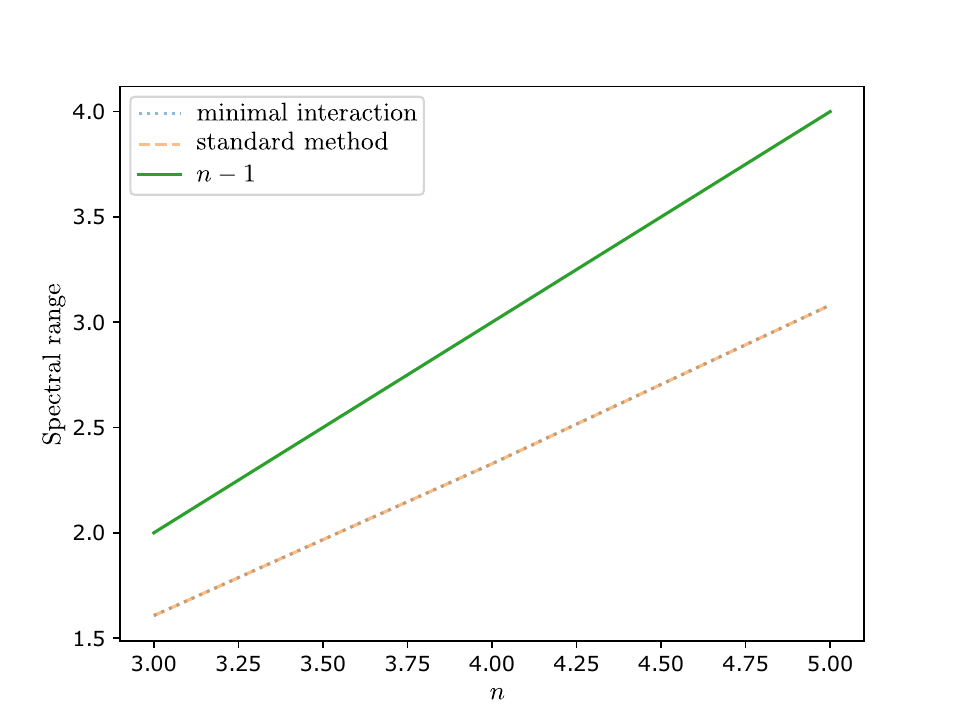}
    \caption{We compute the average spectral range over 1000 randomly sampled initial spectra (each with spectral range \( n - 1 \)) and 1000 randomly sampled target vectors. The results compare the average spectral range achieved using the minimal switching method against that of the standard switching approach.
  }
    \label{fig:min_int}
\end{figure}

While this minimal approach may sacrifice some spectral range, numerical evidence suggests that for small systems, there is no significant performance loss, see Fig.~\ref{fig:min_int}. In this case, the linear system has a unique (up to scaling) solution. To find the best solution, one can vary the scaling factor or introduce additional degrees of freedom by increasing the number of permutation matrices.

However, identifying the optimal sequence of two-permutation matrices involves solving up to
\[
\left( \frac{n(n-1)}{2} \right)^k
\]
linear programs—one for each sequence of permutations—which becomes intractable for large \( n \) due to super-exponential scaling.

For each fixed permutation sequence \( \boldsymbol{\sigma} \), the corresponding linear program can be expressed as
\begin{align*}
    &\textbf{maximize} \quad f_{\boldsymbol{P}}^T v, \\
    &\text{subject to} \quad A_{\boldsymbol{\sigma}, t} v = 0, \\
    &\quad\quad\quad\quad\;\;\; 0 \leq v_i \leq 1,
\end{align*}
where the constraint matrix is defined by
\[
(A_{\boldsymbol{\sigma}, t})_{ij} = \sigma^j(\lambda_i) - (1 - t_i)\sigma^j(\lambda_0) - t_i \sigma^j(\lambda_{n-1}), \quad \text{for } i < n,
\]
and \( A_{n,j} = 1 \). The cost function vector is defined by
\[
f_i = \sum_j \lambda_j (P_{0,j}^{(i)} - P_{n-1,j}^{(i)}).
\]

Due to the super-exponential growth of the search space, this minimal-interaction approach is practical only for small system sizes.

\end{document}